\documentclass[prl,twocolumn,superscriptaddress,showpacs]{revtex4}
\usepackage{amssymb,graphicx}

\begin{document}

\title{Qubits from tight knots and bent nano-bars}

\author{Victor Atanasov}
\altaffiliation[Also at]{ Laboratoire de Physique Th\'{e}orique et
Mod\'{e}lisation , Universit\'{e} de Cergy-Pontoise,
 F-95302 Cergy-Pontoise, France}
\affiliation{ Institute for Nuclear Research and Nuclear Energy,
Bulgarian Academy of Sciences,  72 Tsarigradsko chaussee, 1784
Sofia, Bulgaria}
 \email{victor.atanasov@u-cergy.fr}

\author{Rossen Dandoloff}
\affiliation{ Laboratoire de Physique Th\'{e}orique et
Mod\'{e}lisation , Universit\'{e} de Cergy-Pontoise,
 F-95302 Cergy-Pontoise, France}
 \email{rossen.dandoloff@u-cergy.fr}

\begin{abstract}
We propose a novel mechanism for creating a qubit based on a tight knot, that is a
nano-quantum wire system so small and so cold as to be quantum coherent with respect to curvature-induced effects. To establish
tight knots as legitimate candidates for qubits, we propose an effective
curvature-induced potential that produces the two-level system and identify the tunnel coupling between the two local states. We propose also a different design of
nanomechanical qubit based on twisted nano-rods. We describe how both devices can be manipulated. Also we
outline possible decoherence channels, detection schemes and experimental setups.
\end{abstract}


\maketitle

\begin{figure}[b]
\begin{center}
\includegraphics[scale=0.55]{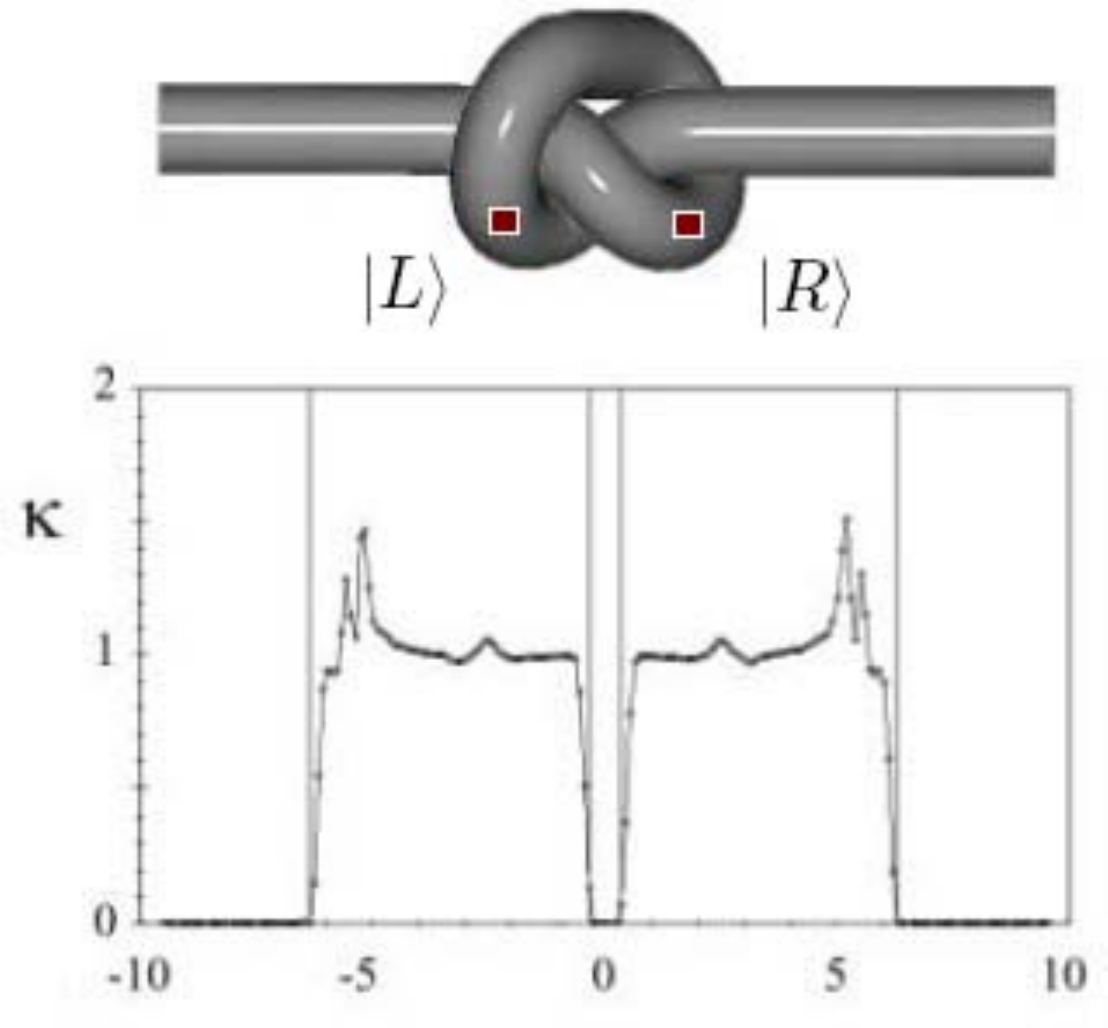}
\caption{\label{fig:knot} The geometry of a trefoil knot and its curvature profile taken from\cite{Pieranski1}. The map of curvature of the tight open trefoil knot shows a mirror symmetry (the mirror plane is vertical and is in the center of the map). On the figure the curvature is normalized with respect to the diameter ($2 \rho_0$) of the tightly knotted thread, that is $(2 \rho_0)^{-1} =1.$  Superimposed on the image is the quantum description suggested in the present study. This description comes in naturally as a result of the exhibited mirror symmetry and the curvature-induced quantum potential.}
\end{center}
\end{figure}

Mathematics views knots as closed, self-avoiding
curves embedded in a three-dimensional space\cite{Kauffman}, e.g. any knot can be tied on
a thread of any length. This definition leads to the conclusion that a knot tied on a  thread can be arbitrarily shaped and all its
conformations are essentially equivalent. Besides, the lack of characteristic  length prevents the introduction of energy scale. Physics views real knots differently\cite{Ashley} since tying them
requires the use of thread of a finite diameter and tying a
particular type of knot requires a thread piece with
proper length. In the physically relevant situation the diameter will play the role of characteristic length.

Suppose we are to tie a knot on a a thread with circular cross-section and pull the knot tight. 
This process has its limit. There exists a particular
conformation of the knot at which the thread's ends
cannot be pulled apart any more without destroying it. The final conformation will be referred to as {\it tight open knot}. If the loose ends are connected continuously we will refer to the resulting structure as {\it tight closed knot}. An argument based on the minimization of the elastic energy in the material on which the knot is tied points to the {\it symmetrical} conformation as most energetically favorable. Thus the existence of a plane of symmetry both for the open and the close tight knot is suggested (see Figure \ref{fig:knot}). 

As indicated by de Gennes\cite{Gennes}, such knots are spontaneously
tied and untied by thermal fluctuations on long
polymeric molecules and change
the macroscopic properties of materials\cite{Grossberg} which is one of the very few practical applications of knot theory in physics. In this paper we attempt to extend the scope of practical applications of knot theory and the idea that a collection of nano-rods glued together can produce prescribed quantum behavior for a particle trapped inside\cite{vic*07}. 

The thread's cross sections are shaped as disks of radius $\rho_0$ which centers are located on a centerline whose tangent $\vec{t}$ is  continuous (no cusps). Their normals coincide with the tangent $\vec{t}$ at every point. The 
disks are not allowed to overlap which is the self-avoiding condition. The effect of the self-avoiding
condition on the curvature $\kappa$ of the thread's centerline is the following $ \rho_0^{-1} \kappa < 1.$

In what follows the thread represents a quasi-one-dimensional quantum wire following the centerline and dressed with an insulating material up to its diameter $2 \rho_0$.

The curvature profile of the unique
tightest conformation of the trefoil open knot tied on a perfect thread is discussed within the numeric experiment in\cite{Pieranski1}. The result is conveyed in Figure \ref{fig:knot}.

We will consider possible quantum mechanical implications of this profile
for $\kappa$, with regard to electron transport and formation of bound states on the quantum wire. It has been shown by \cite{daCosta1} that a
quantum particle on a quantum wire whose axis follows a space curve with curvature $\kappa$ feels an effective potential\cite{daCosta1} of the form
\begin{equation}\label{daCosta}
V_{eff}(s)=-\frac{\hbar^2}{2 m}   \frac{\kappa^2(s)}{4}
\end{equation}
where $m$ is the effective particle's mass,
$\hbar$--Plank's constant and $s$ is the arclength. 

\begin{figure}[ht]
\begin{center}
\begin{picture}(150,80)

\put(5,60){\vector(1,0){145}}
\put(70,5){\vector(0,1){80}}
\put(70,60){\circle*{2}}
\put(63,63){0}
\put(140,63){s}

\put(1,48){$0$}
\put(7,50){\line(1,0){29}}
\put(7,50.5){\line(1,0){29}}

\put(65,50){\line(1,0){10}}
\put(65,50.5){\line(1,0){10}}

\put(65,50){\line(0,-1){40}}
\put(65.5,50){\line(0,-1){40}}

\put(35,50){\line(0,-1){40}}
\put(35.5,50){\line(0,-1){40}}

\put(35,10){\line(1,0){30}}
\put(35,10.5){\line(1,0){30}}

\put(45,1){$| L \rangle$}
\put(85,1){$| R \rangle$}

\put(75,50){\line(0,-1){40}}
\put(75.5,50){\line(0,-1){40}}

\put(105,50){\line(0,-1){40}}
\put(105.5,50){\line(0,-1){40}}

\put(75,10){\line(1,0){30}}
\put(75,10.5){\line(1,0){30}}

\put(120,10){\line(1,0){10}}
\put(125,14){\vector(0,1){36}}
\put(125,50){\vector(0,-1){39}}

\put(133,25){$k_0^2=\frac14 \kappa^2$}

\put(75,35){\line(1,0){30}}
\put(80,50){\line(1,0){10}}

\put(80,14){\vector(0,1){20}}
\put(80,35){\vector(0,-1){25}}

\put(82,20){$k^2$}

\put(85,35){\vector(0,1){15}}
\put(85,50){\vector(0,-1){15}}

\put(88,40){$q^2$}

\put(35,30){\line(1,0){30}}
\put(35,40){\line(1,0){30}}

\put(40,30){\line(0,1){20}}
\put(40,40){\line(0,-1){20}}

\put(40,25){\vector(0,1){5}}
\put(40,45){\vector(0,-1){5}}

\put(43,45){\footnotesize{$\Delta E$}}

\put(105,50){\line(1,0){40}}
\put(105,50.5){\line(1,0){40}}

\end{picture}
\end{center}
\caption{\label{fig1}The curvature-induced quantum potential for the knot's centerline.}
\end{figure}
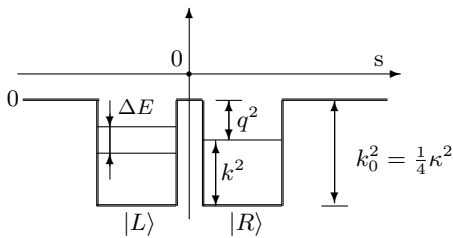

Other study\cite{Neukirch} of elastic rods bent into open loose knots within the Kirchhoff equations for rods report similar symmetric profile for the curvature with respect to the point of symmetry of the knot. Thus we believe that the simplified potential that we adopt and is depicted in Figure \ref{fig1} is justified in view of the numerical data\cite{Pieranski1} and the analytic solution\cite{Neukirch}.

The problem of quantum dynamics for an electron in the presence of an idealized curvature-induced quantum potential of binding quality depicted in Figure \ref{fig1} leads us to the striking observation that due to the symmetry of the trefoil knot a two-well potential along the arclength of the knot's central curve emerges.

A particle once prepared in a state occupying only one of the wells with certain energy $E$, that is $| L\rangle$ or $| R \rangle$ state, possesses tunneling probability to filter through the barrier between the two wells along the arclength of the knot's central curve. We can construct an approximate solution in this energy field due to curvature using the quasi-classical wave function $\psi(s)$ describing the motion with energy $E$ in one of the wells and exponentially decreasing on both sides of the well's boundaries.  Using this approximate solution and a model potential well of depth $U_0={\hbar^2 \kappa^2}/{8 m} = {\hbar^2 }/{(2^5 m \rho_0^2)} $ and width $D = 5 \rho_0$ (where we have taken model's values from observation of Figure \ref{fig:knot}) we can quantify the energy split due to tunneling.

The bound state energies $E=-\frac{\hbar^2}{2m}q^2$ corresponding to such curvature-induced effective potential are given by
\[
E_n=-U_0 \left[ 1 - \left[  {(k_n D)}/{(2 C)} \right]^2    \right],
\]
where $U_0=\hbar^2 k_0^2 / 2 m$ is the depth of the well in terms of  $k_0^2=\kappa^2 /4$ for the knot, $D$ is the width of the well, that is the length of the stretched section where the quantum particle relaxes due to Heisenberg's uncertainty principle, and $C={\kappa D}/{4}$ is a geometrically determined quantity qualifying the potential well. According to Figure \ref{fig:knot} we can take $D=5 \rho_0$ and $\kappa=1/(2 \rho_0)$ which fixes $C=5/8.$ Here the discrete values $k_n$ of the wave number $k^2=k_0^2-q^2$ are solutions to the algebraic equations
\begin{equation} \label{eq for k}
\tan{k \frac{D}{2} } = \frac{\sqrt{C^2 - \left( k \frac{D}{2} \right)^2  } }{ k \frac{D}{2} } \left(   = -\frac{ k \frac{D}{2} }{\sqrt{C^2 - \left( k \frac{D}{2} \right)^2  } }
     \right)
\end{equation}
for the even (and  for the odd) eigenstates. It is interesting to notice that {\it the existence of at least one even ground state} is guaranteed. It is  
\begin{equation}\label{eq:k even/odd}
k_1 \approx \frac{1}{5 \rho_0}, \qquad E= -\left(\frac{3}{5} \right)^2 U_0 \; \approx -\frac{3^2\hbar^2 }{2^5 5^2 m \rho_0^2} .
\end{equation}
With the above value $C=5/8$ we have only { \it one even state} (\ref{eq:k even/odd}). In practice, changing knot's conformations we may adjust the value of $C$ and generate more levels.

It is interesting to point out the situation in which the quantum wire on which the knot is tied has a finite length with an infinite potential barrier at both ends. This leads to the condition of a node at the ends and modifies (\ref{eq for k}) which for a knot tied in the middle of the rope's extent is
\begin{equation}\label{eq:k solid}
\nonumber k D = \sum_{\epsilon=0}^{1} \tan^{-1}{\left\{ \pm \left[\frac{q}{k} \coth{q\left(l + (-1)^{\epsilon} \frac{D}{2} \right)} \right]^{\pm 1} \right\} } 
\end{equation}
for the even ($+$) and
for the odd ($-$) eigenstates. Here $l$ measures the distance from the knot to the either end. The boundary conditions do not affect the existence of the {\it guarantied} negative energy even eigenstate for which we can again use (\ref{eq:k even/odd}). The hard walls introduce quantization in the positive energy spectrum up to a finite number of positive energy eigenstates. The second even eigenstate after (\ref{eq:k even/odd}) is at $k_2=7/(25 \rho_0)$ for $l=5D$ which corresponds to energy $E_2 = 3 U_0 /25.$ The first odd eigenstate appears at $k_1^{odd}=4/(25 \rho_0)$ or $E_1^{odd} = 7 U_0 /25.$ As $l$ is increased they shift slightly and for $l \to \infty$ we are left with (\ref{eq:k even/odd}).

An estimate of the number of states $N_s$ as a function of the geometry and the topology of the knot can be obtained evaluating the integral\cite{Landau}
\begin{equation}
N_s=\int \frac{\sqrt{-2 m V_{eff}(s)} d s}{2 \pi \hbar} = \frac{1}{4 \pi} \int \kappa(s) ds.
\end{equation}
Now let us consider a {\it tight closed knot}. A property of knots on closed curves\cite{Fary} is the existence of a lower bound for $ \oint \kappa(s) ds \geq 4 \pi.$ Thus $N_s \geq 1$ and a bound state in the case of a tight knot tied on a closed nano wire does exist! Due to the symmetry of the tightest conformation it is again split into two levels $\Delta E$ apart.

The knot becomes a two-level quantum mechanical system or in other words a qubit,  due to tunneling. The energy level $E$ splits into two $E^+$ and $E^- .$ The zeroth-approximation wave functions corresponding to the two levels are symmetrical and anti-symmetrical combination of the quasi-classical wave function  $\psi(s)$ and $\psi(-s),$ that is $ 
\psi^{\pm} (s)=\left[  \psi(s) \pm \psi(-s)  \right]/\sqrt{2}.
$
The split is expressed as
$
\Delta E=E^+ - E^-=\frac{2 \hbar^2}{m} \psi(0) \frac{d \psi}{d s} (0), 
$ where the zero is chosen so as the axis of symmetry of the potential passes through (see Figure \ref{fig1}) it and the quasi-classical wave function is given in the standard way\cite{Landau}. The width of the energy split is
\begin{equation}\label{eq:DeltaE}
\Delta E=\frac{\Omega \hbar}{\pi}  e^{ - \frac{1}{\hbar} \int_{-a}^{a} \left| p_1  \right| d s  }  =\frac{ \hbar^2 | k_1 |}{m D} e^{ - | k_1 | d },
\end{equation}
where  the order of magnitude is $\frac{ \hbar^2 }{5^2 m \rho_0^2 }e^{ - d/(5\rho_0) }$ and $\Omega$ is the frequency of the classical periodic motion in the well. It is given in terms of the period for the classical motion
\[
\Omega=\frac{2 \pi}{T}=\frac{\pi}{m} \left(  \int_{b}^{a} \frac{d s} {p_n}  \right)^{-1}=\frac{\pi \hbar | k_1 |}{m D} \sim \frac{\pi \hbar }{5^2 m \rho_0^2},
\]
where $a$ and $b$ are the turning points corresponding to the periodic trajectory in the well, $D=|a-b|$ is the width of the potential well, the associated momentum $p_1$ is $
p_1=\sqrt{2 m (U_0-E)}=\hbar | k_1 | \sim {\hbar}/{(5 \rho_0)}.$
Here $d$ is the width of the barrier between the two wells. 

All quantities in (\ref{eq:DeltaE}) are geometry determined by the knot's conformation!

\begin{figure}[b]
\begin{center}
\includegraphics[scale=0.10]{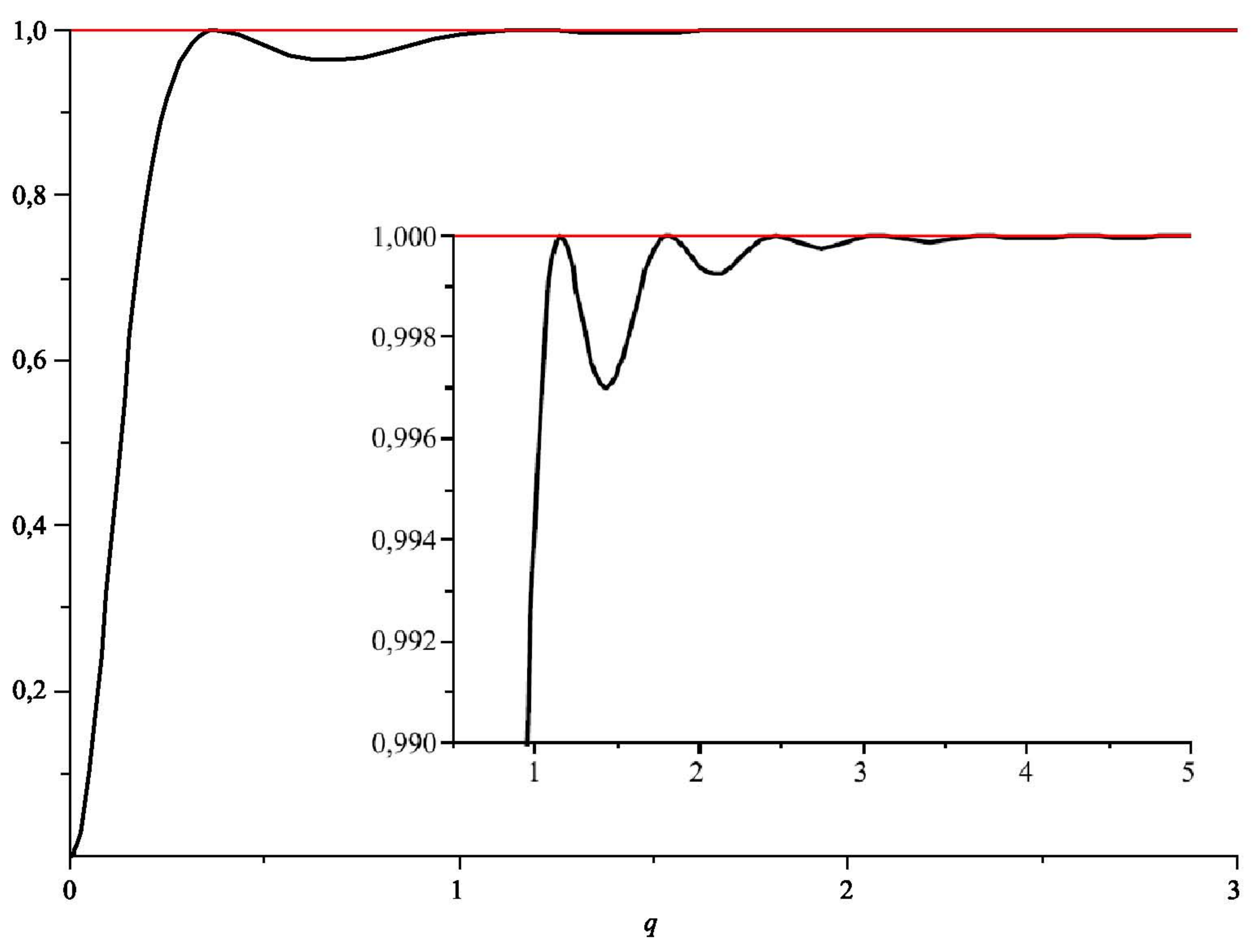}
\caption{\label{fig:flux} The ratio between the transmitted and the incident probability currents ${j_t}/{j_i}$ as a function of the wave number $q=\sqrt{2 m E}/\hbar$. Here we have set $D=5/2,$ $\rho_0=1/2,$ $\kappa=1$ and $d=10^{-2}.$}
\end{center}
\end{figure}

Now if we propagate a plane wave with positive energy from one end of the knot to the other we may acquire insight into the properties of the curvature-induced potential by measuring the ratio between the transmitted $j_t$ and incident $j_i$ currents. For the model potential (see Figure \ref{fig1}) theory gives a lengthy expression\cite{tunneling}. Conveyed graphically in Figure \ref{fig:flux} is the behavior of $j_t/j_i.$ The figure renders visible the
energies for which the knot is completely transparent. The first few energies for which ${j_t}/{j_i}=1$ with the data from Figure \ref{fig:flux} are summarized in the following table:

\begin{center}
\begin{tabular}{|c|c|c|c|c|c|c|}
\hline $i$ & 1 & 2 & 3 & 4 & 5 & 6 \\\hline $q_i=\sqrt{2m E_i}/\hbar$ & 0.38 & 1.15 & 1.81 & 2.46 & 3.06 & 3.74 \\\hline 
\end{tabular} 
\end{center}

Having identified the energy level and its split both fixed by the knot's geometry, let us turn our attention to the possible effects that internal (along the wire) and external electric fields can have on the operation of such a device. We also assume that the application of such fields does not influence the geometry of the knot.

The application of an electric field along the thread would lead to a shift in the potential energy making the well asymmetrical. The potential difference between the two ends of the well will be $\Delta V= e \mathcal{E}D.$ Here $\mathcal{E}$ is the field's strength, $e = \mp |e|$ is the charge of the particle assumed to be an electron or a hole respectively. The critical value at which the level disappears is to be found by the standard inequality\cite{Landau}
$
\cos{\frac{\kappa  D}{2}} \geq - \sqrt{1+ \frac{8m}{\hbar^2 \kappa^2} e \mathcal{E} D }
$
which for weak field leads to an estimate for the non-destructive field strength
\begin{equation} 
0 \leq \mathcal{E} \leq  \frac{\hbar^2 \kappa^2} {4m e D} \left(1 +\cos{D \frac{\kappa}{2}} \right) \leq \frac{\hbar^2 } {20 m e \rho_0^2}.  
\end{equation}

The entire knot can be placed in an external field that could be time dependent  $\mathcal{E}_{ext}(t).$ If we place a charged particle in one of the wells, say $| L \rangle$ or $| R \rangle$ state, the knot becomes electrically asymmetrical and an effective electric dipole moment $\mu=e(d + D)$ can be introduced. External fields can couple with it thus reminisce the way ammonia molecule is driven by external electric field at resonance frequency. For the knot this resonance frequency is calculated to be
\begin{equation}\label{omega}
\omega=\frac{\Delta E}{\hbar}=\frac{\hbar | k_n |}{m D} e^{ - | k_n | d } \sim \frac{ \hbar }{5^2 m \rho_0^2 }e^{ - d/(5 \rho_0) }.
\end{equation}
The knot qubit can be prepared in the $| R \rangle$ state by setting a longitudinal electric field with respect to Figure \ref{fig:knot}  towards the right (assuming the particle in the interior of the knot is negatively
charged, say an electron). By suddenly turning off this electric field the knot's state is prepared in a coherent superposition of $(| L \rangle \pm | R \rangle)/\sqrt{2}.$ Because of
the tunnel splitting, the system then starts to oscillate coherently, with a frequency given by (\ref{omega}).

After an amount of time $t,$ the knot can be in either $| L \rangle$ or $| R \rangle$ states. Therefore, by detecting
the particle's distribution, as a function of $t$, we can determine the coherent oscillation
frequency experimentally. Driven transitions between the two states
can be achieved by adding a sinusoidal component at the resonance frequency.

Quantum computing requires two-qubit operations. For charged knots, the
inter-qubit interaction comes naturally in terms of the electric dipole interaction due to $\mu$ between the knots. The application of  longitudinal electric field can be used to tune selected knots into resonance, then  microwaves can be used to perform conditional rotations and other operations.

Qubits can operate only if decoherence is low enough. For quantum wires major sources of noise are the internal thermomechanical noise (due to non homogeneous
distribution of elastic energy in the knot's interior), noise from moving defects in the material medium, etc.
In the quantum mechanical limit, some of these noises become unimportant. There
should be no heat flow.  The whole system has to be cooled down to
\[
T < \frac{U_0}{k_B} \sim 1K \; (\rho_0 \sim {\rm nm})
\]
and put in vacuum in order to reach the two lowest states of the knot. The main source
of quantum mechanical decoherence might be the internal dissipation caused by electron--phonon
interactions in the material medium. Since the knots are to be attached on to a much larger system, coupling at the ends of the knot could lead to relaxation/excitation
of the qubit states. Another source of decoherence could be charge fluctuations. All of these could lead to the loss of quantum interference.

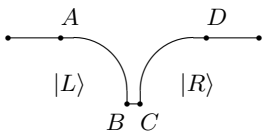
\begin{figure}[ht]
\begin{center}
\begin{picture}(150,50)

\put(65,10){\circle*{2}}
\put(57,0){$B$}

\put(70,10){\circle*{2}}
\put(70,0){$C$}

\put(65,10){\line(1,0){5}}

\put(40,10){\oval(50,50)[tr]}
\put(95,10){\oval(50,50)[tl]}

\put(40,35){\circle*{2}}
\put(40,40){$A$}

\put(37,15){$| L \rangle$}

\put(95,35){\circle*{2}}
\put(95,40){$D$}

\put(85,15){$| R \rangle$}

\put(95,35){\line(1,0){20}}
\put(115,35){\circle*{2}}

\put(40,35){\line(-1,0){20}}
\put(20,35){\circle*{2}}

\end{picture}
\end{center}
\caption{\label{fig4}A collection of circular and straight nanobars which recreates the double-well curvature-induced potential due to confinement as depicted in Figure \ref{fig1}.}
\end{figure}

Now let us turn our attention to the description of the device depicted in Figure \ref{fig4}. It is made up of a collection of straight and bent (into quarter circles with radius $R$) nano-rods. The quantum effective potential within the glued rods is exactly the same in form as the one for the knot (Figure \ref{fig1})  with $D=\pi R /2.$ Here $d=||BC||$ and $D$ can be adjusted by picking rods with appropriate (with respect to our resolving power in measuring $E,$ since the energy split is $\Delta E \approx D^{-2} \exp{[-d/D]}$) lengths and radius of curvature. Due to the similarity in the quantum potential this device can be manipulated in exactly the same manner as the knot qubit.

\begin{figure}[b]
\begin{center}
\begin{picture}(150,90)

\put(20, 50){\oval(5,80)}
\put(110, 50){\oval(5,80)}

\put(13,75){\vector(1,0){120}}
\put(123,80){s}
\put(60,80){s=0}

\put(25,78){\footnotesize{E=0}}

\put(65,75){\circle*{2}}

\put(65,55){\line(-1,-1){43}}
\put(65,55){\line(1,-1){43}}

\put(25,55){\line(1,0){10}}
\put(30,57){\vector(0,1){17}}
\put(30,75){\vector(0,-1){19}}

\put(33,62){\footnotesize{$k_0^2=\frac{\kappa^2}{4}$}}

\put(85,35){\line(1,0){23}}
\put(90,30){\line(1,0){17}}
\put(95,25){\line(1,0){13}}

\end{picture}
\end{center}
\caption{\label{fig5} The quantum potential within the nano-bars' interior in the presence of suitably applied electric  field $\mathcal{E}$.}
\end{figure}
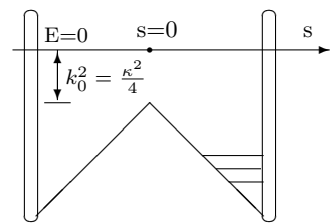

It is interesting to note also that if we choose $d=0,$ that is we remove the central rod and actually merge the points $B$ and $C,$ we can produce another device which can serve as a qubit. By applying symmetrically electric field with respect to the point of merger (also a point of symmetry), say ($-$), and points $A$ and $D,$ say ($+$), in the manner depicted in Figure  \ref{fig5} we can create controllable energy split of the quasi-stationary states in the well created by the electric field and the insulator at the end points on the figure.  Such a device is vulnerable to the Nyquist-Johnson noise from the driving electrical generator.

In conclusion we have expanded the application of knot theory in physics by recognizing that a tight knot tied on a quantum wire with nanoscale diameter creates in its interior a quantum potential due to curvature of binding quality. An energy level in this potential is split into two due to the symmetry of the knot, a property resulting from the minimization of the elastic energy in the material on which it is tied. Similar potential can also be created by a collection of bent nano-rods glued together. Electric fields can be applied to manipulate and tune such a system due to its electrical asymmetry if a charged particle is occupying the left or the right well of the double-well potential producing the two-level system. This electrical asymmetry also couples identical knots and serves as inter-qubit interaction. All of this suggests experimental setups which could be realized with present technology.

\end{document}